\documentclass[a4paper]{article}

\usepackage{atmohead2013}
\usepackage[english]{babel}

\title{Early Attempts at Active Atmospheric Calibration with H.E.S.S. Phase 1}

\shorttitle{Active Atmospheric Calibration for H.E.S.S. Phase 1}

\authors{
Nolan, S.J.$^{1}$,
Rulten, C.B.$^{1,2}$,
P\"uhlhofer, G.$^{3}$
}

\afiliations{
$^1$ Physics Department, Durham University, South Road, Durham, County Durham, DH1 3LE \\
$^{2}$ now at: LUTH, L'Observatorie de Paris, 5, place Jules Janssen, 92195 Meudon, FRANCE \\
$^{3}$Landessternwarte, Universit\"at Heidelberg, K\"onigstuhl, D 69117 Heidelberg, Germany\\
}

\email{s.j.nolan@dur.ac.uk}

\abstract{Using data derived from the H.E.S.S. Phase 1 telescope system and a Ceilometer facility on site, a method of correcting for changing atmospheric quality based on reconstructed shower parameters is presented. The method was applied to data from the active galactic nucleus PKS 2155-304, taken during August and September 2004 when the quality of the atmosphere at  the site was highly variable. Corrected and uncorrected fluxes are shown, and the method is discussed as a first step towards a more complete atmospheric calibration.}

\begin{document}
\maketitle
\section{Introduction}Imaging Atmospheric Cherenkov Telescopes (IACTs) rely heavily on the atmosphere as their detecting medium. Although the atmosphere gives the telescope systems huge effective areas, daily variations in atmospheric quality can affect the system performance and lead, in the worst cases, to systematic bias in the estimated energy of a given event. Significant effort has been made in the past to take account of this problem by using the cosmic-ray background seen by the telescope on a given night to normalise the data \cite{bib:bohec}. However, given a better understanding of the location of atmospheric aerosol populations from LIDAR measurements and via modelling of these populations, it is possible to determine an active atmospheric correction to the data. Herein, recent work on such a technique is discussed as applied to observations with the H.E.S.S. telescope array of the active galactic nucleus (AGN) PKS 2155-304.

\section{Technique}
\begin{figure*}[th]
\begin{center}
\includegraphics*[width=0.73\textwidth,height=0.30\textheight,clip]{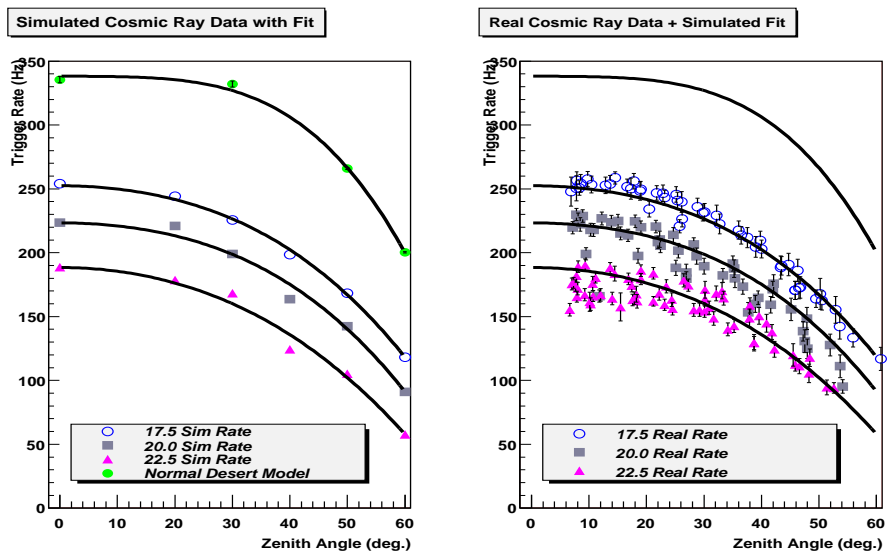}
\caption{\label {fig1} Simulated array trigger-rate for a spectrum of cosmic-rays \cite{bib:karle} for various atmospheric models with function fit in left panel, versus measured cosmic-ray trigger rates for the PKS 2155-304 2004 dataset in right hand panel. }
\end{center}
\end{figure*}

\begin{figure*}[th]
\begin{center}
\includegraphics*[width=0.80\textwidth,height=0.24\textheight,clip]{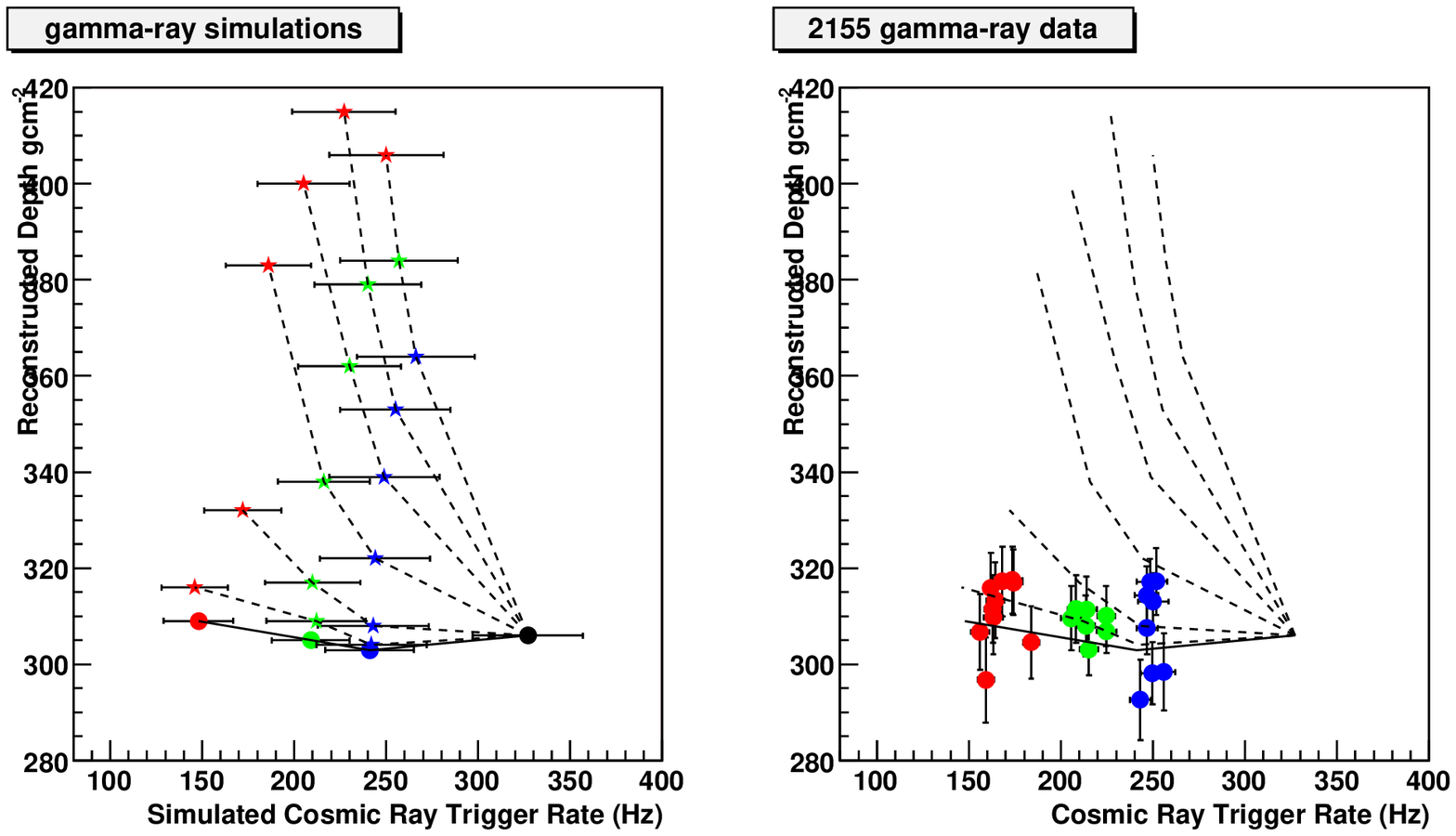}
\caption{\label {fig2} The left panel shows the mean of reconstructed depth (for a Gaussian fit) for gamma-ray shower simulations at 20 degrees zenith-angle versus telescope trigger-rate. The lower points (solid circles) show the results for the 17.5, 20.0 and 22.5 m/s wind speed models, with the other points showing show the result for atmospheres with increasing altitude of the aerosol contaminant layer, with lines connecting similar altitudes. These lines are reproduced on the right hand plot, which shows the preliminary real mean reconstructed depth for gamma-ray data on PKS 2155-304 taken during 2004 at zenith angles between 15 and 25 degrees, slightly scaled to match the results at 20 degrees.  The data shown no indication of high level aerosols, independently confirming the Ceilometer results.} 
\end{center}
\end{figure*}

The Ceilometer (simple LIDAR) system based at the H.E.S.S. site between 2003-2007 works at a wavelength of 905 nm, and has an active range of 7.5 km. It is mounted on an alt-azimuth drive allowing on-source pointing during observations. The return signal from the Ceilometer is proportional to the backscattered light  intensity produced through Rayleigh and Mie scattering. In this study the data from the Ceilometer is used to identify the relative night to night aerosol density and the location of the aerosol populations up to an altitude of 7km.  During August and September 2004, a large population of  aerosols was seen by the Ceilometer below 2 km above the site, concurrent with a significant drop in the H.E.S.S. array trigger-rate for cosmic-rays. This population was seen to vary on a night to night basis, but not within a given night. In order to simulate its effects, the atmospheric simulation code MODTRAN was used to generate optical depth tables for wavelengths in the range 200 to 750 nm and for successive heights above the site (which is 1.8 km above sea level) \cite{bib:Modtran}. The aerosol desert model within MODTRAN introduces a homogeneous layer of aerosols into the first 2 km above ground level, whose concentration is then increased as the wind speed parameter is increased. Thus optical depth tables were produced for the range of wind speeds from 0 m/s to 30 m/s. The wind speed therefore acts as a tuning parameter to match simultaneously cosmic-ray trigger-rate and image parameter distributions, and is not a reflection of the measured wind speed at the site. These tables were then applied to a set of CORSIKA cosmic-ray simulations at various zenith angles between 0 and 60 degrees and with a southern pointing, to best match the data taken on PKS 2155-304, and a cosmic-ray trigger-rate for each atmosphere was derived for the H.E.S.S. array based upon the spectra given in \cite{bib:karle}. By matching the trigger-rate from simulations and real data, taking into account zenith angle dependence effects and gain changes over the experiment lifetime, an atmospheric model can be selected, as discussed in \cite{bib:nolan}. The real cosmic-ray trigger rate and that due to simulation for the PKS 2155-304 dataset discussed later are shown in figure \ref{fig1} for comparison. The figure clearly shows that the data can be separated into 3 classes corresponding to MODTRAN model wind speeds of 17.5, 20.0 and 22.5 m/s. \newline
 In addition, as the Ceilometer has a limited range and sensitivity, and to further confirm the choice of atmospheric models, a set of atmospheric models with aerosol densities at higher altitudes was simulated using MODTRAN. These simulated atmospheres represent conditions which could in principle also have occured during data-taking, as they result in similar cosmic-ray trigger rates as the low-level aerosol models.  As shown in figure \ref{fig2}, by comparing the reconstructed shower depth for gamma-rays between real-data and simulations, these models are shown to be considerably less favoured than the simple low-level aerosol models of 17.5, 20.0 and 22.5 m/s wind speed, which trigger-rate, image parameters, mean shower-depth and Ceilometer data validate. \newline
The atmospheric model is then applied to a full set of 18 million CORSIKA gamma-ray simulations  within a telescope simulation code. The simulations cover the zenith angle range of the observations, and produce lookup tables for image parameter cuts, energy and effective area, and these in turn are applied to the data using the standard H.E.S.S. analysis procedure \cite{bib:hess2}. 
\section{PKS 2155-304}PKS 2155-304 is an AGN of the blazar class at a redshift of z$=0.116$. It was first detected in TeV gamma-rays by the Durham Mark 6 telescope \cite{bib:chadwick}, and has been observed from the earliest days of the H.E.S.S. experiment \cite{bib:hess1}.   The data set from August and September 2004 is formed from 86 hours of four telescope observations. The exposures of this dataset under different atmospheric conditions is shown in \ref{tab1}. 
\begin{table}[tbh]
\begin{center}
{\footnotesize
\begin{tabular}{| c | c | c |}
\hline
Wind Speed&Exposure (Hours)& $\%$ of \\
& &Total Exposure\\
\hline
17.5&23.7&27.4\\
20.0&19.5&22.6\\
22.5&43.4&50.0\\
\hline
\end{tabular}
}
\end{center}
\caption{Table showing the exposures of the AGN PKS 2155-304 as a function of atmospheric conditions during August 2004.}   
\label{tab1}
\end{table}

\begin{figure*}[th]
\begin{center}
\includegraphics[width=0.95\textwidth,height=0.3\textheight,clip]{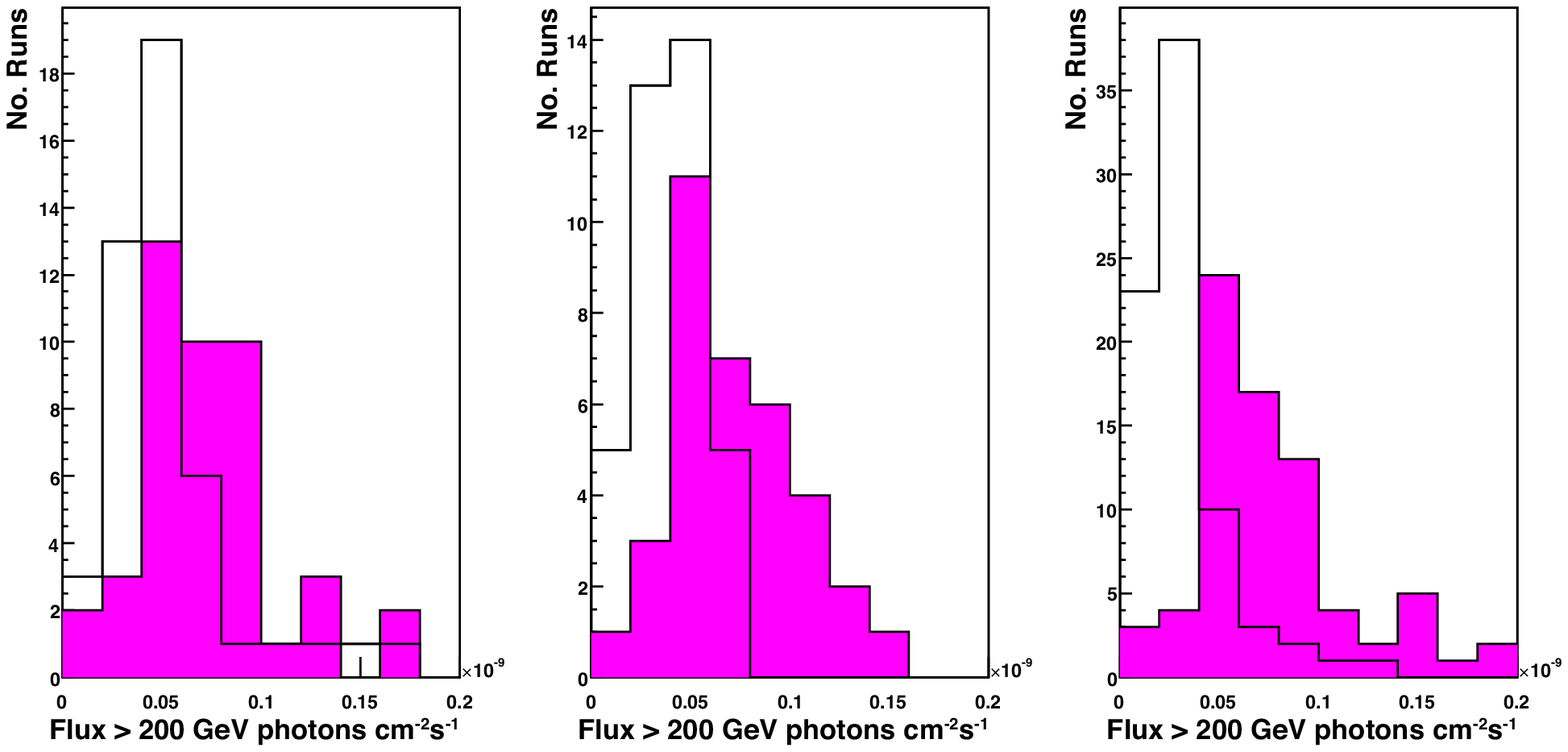}
\caption{\label{fig 3} The preliminary distribution of muon corrected integral flux for PKS 
2155-304 above 200 GeV derived from 28 minute runs is plotted before (open histograms) and 
after (filled histograms) the application of corrections for low-level dust. As noted each panel shows a subset of the data of differing atmospheric class, with left panel showing the atmospheric class with a wind speed of 17.5 m/s, the centre panel showing that for a wind speed of 20.0 m/s and the right panel showing the flux distribution for a wind speed of 22.5 m/s.}
\end{center}
\end{figure*}

\begin{figure*}[hb]
\begin{center}
\includegraphics[width=0.95\textwidth,height=0.30\textheight,clip]{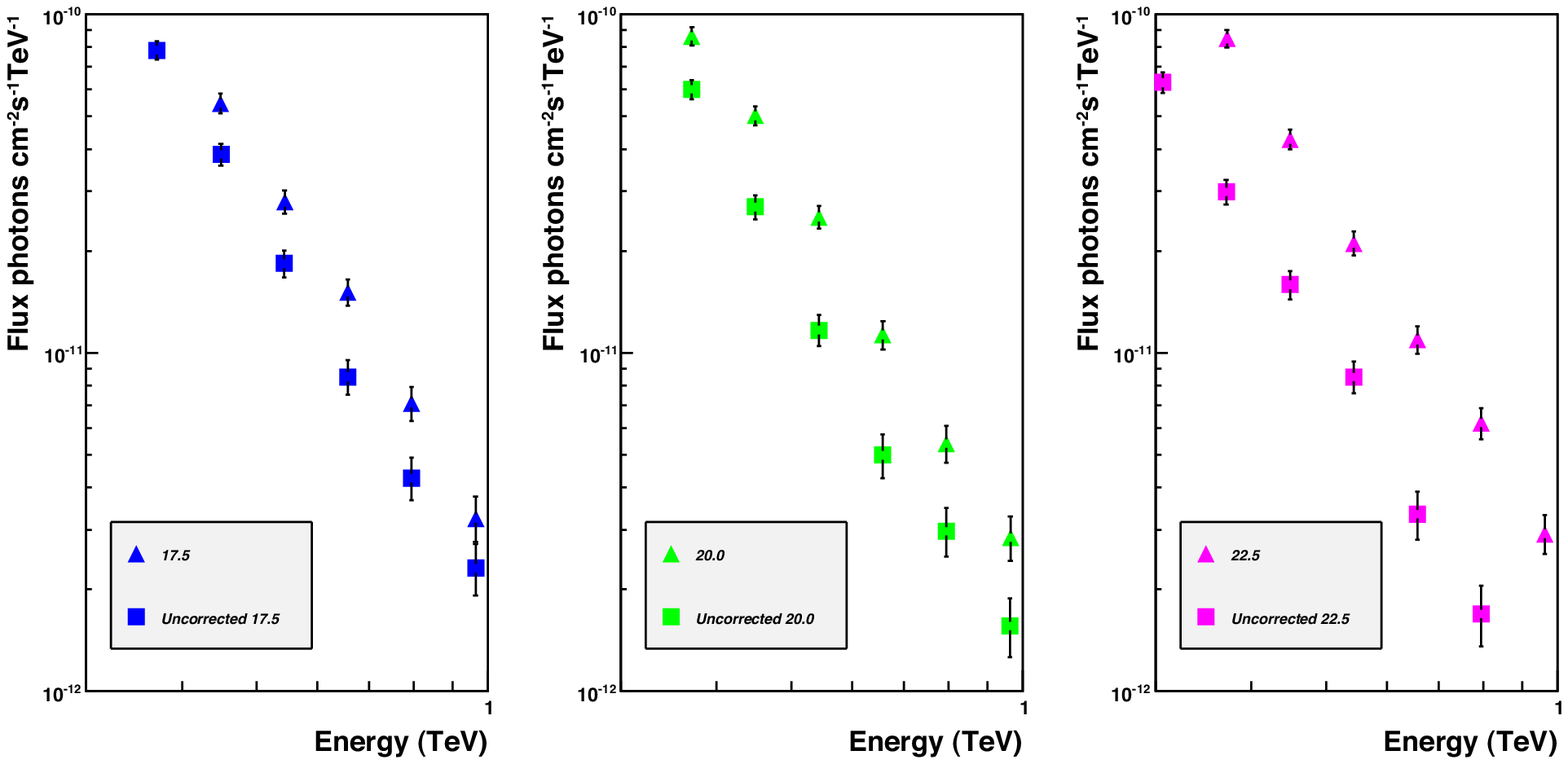}
\caption{\label{fig4}The preliminary uncorrected and corrected differential spectral for the 3 subsets of data is shown between 300 GeV and 1 TeV.  Above 1 TeV differences are negligible compared to statistical errors.  }
\end{center}
\end{figure*}

By combining flux data into atmospheric correction groups, figure \ref{fig 3} shows the results for corrected and non-corrected data in the form of a plot of the flux distribution derived on a run by run basis.  It appears that in the data set considered here, as no run was taken under normal, clear atmospheric conditions, all runs are subject to systematically lowered detection rates, which if uncorrected may lead to significantly different results.
In addition, figure \ref{fig4} shows the spectra derived from these data. Without correction, significantly different results are arrived at, with spectral index for a power-law fit differing by (at worst) $\Delta=0.7$, which is within errors marginally incompatible with a constant index. With correction all fit spectral indices agree well within errors.   

We are confident that the presented method yields the correct energy spectra for PKS 2155-304. Given however the variable nature of the source, neither flux nor photon index is expected to match results obtained from observations in other years, therefore the results cannot be verified by such comparisons. A constant gamma-ray source of sufficient brightness that could be used as calibrator was not observed with H.E.S.S. during the aerosol contamination period. We were however able to identify a small set of Crab Nebula data that were taken under similar atmospheric conditions, despite the fact that Crab observing season is in the other half of the year. Indeed, those ~1.5 hours of observations can best be described -- according to cosmic-ray trigger rate and ceilometer return signal -- by the 17.5 m/s wind speed parameter atmospheric model. The parameters for a power-law fit to the differential spectrum of this Crab Nebula data are shown in Table \ref{tab2}, along with the fit parameters for the PKS 2155-304 dataset. The application of the appropriate reconstruction to the Crab data leads to a very significant increase in flux normalisation ($\Delta$ I$_0$ $\simeq $25$\%$), which after correction is in perfect agreement with the published values derived from unaffected data \cite{bib:hess2}. For the photon index $\Gamma$, unfortunately, a meaningful statement is not possible, because of the statistical error and the smallness of the correction ($\Delta \Gamma \simeq$ 0.03). Both uncorrected and corrected values for $\Gamma$ are within errors in agreement with the published value.

\noindent The differences in flux normalisation and photon index between uncorrected and corrected numbers for the PKS 2155-304 data are larger than those for the Crab data. The difference in flux normalisation for PKS 2155-304 shows the expected strong correlation with atmospheric attenuation, $\Delta I_0 \sim$ 38$\%$ (17.5 m/s) - 64$\%$ (22.5 m/s), while $\Delta \Gamma \simeq$ 0.13 for all three subsets. In general, we believe that the different corrections for the Crab and PKS 2155-304 data sets are related to the different photon indices of those sources, since the corresponding reconstructed energy effective areas  when weighted by the different photon indices differ. 
\noindent Overall, it is fairly reassuring that the magnitude of the photon index correction is in the same range as the systematic error quoted for H.E.S.S. photon indices ($\Delta \Gamma \sim$ 0.1), which was estimated for data taken under normal good weather conditions.

\begin{table*}[b]
\begin{center}
{\footnotesize
\begin{tabular}{| c | c | c | c | c | c | c |}
\hline
Source & Type & Atm & Lookup & I$_{0}\times 10^{-12}$ & $\Gamma$ & $\chi^{2}$/NDF \\
 & Model & Used & & & & \\
\hline
Crab Nebula & Uncorrected & 17.5 & std & 26.0$\pm$1.0 & 2.62$\pm$0.07 & 0.67 \\
Crab Nebula & Corrected & 17.5 & 17.5 & 35.0$\pm$2.0 & 2.59$\pm$0.07 & 0.77 \\
\hline
\hline
PKS 2155-304 & Uncorrected & 17.5 & std & 2.1$\pm$0.2 & 3.50$\pm$0.07 & 2.3 \\
PKS 2155-304 & Corrected & 17.5 & 17.5 & 3.4$\pm$0.3 & 3.38$\pm$0.07 & 2.3 \\
\hline
PKS 2155-304 & Uncorrected & 20.0 & std & 1.4$\pm$0.1 & 3.63$\pm$0.05 & 1.8 \\
PKS 2155-304 & Corrected & 20.0 & 20.0 & 2.7$\pm$0.1 & 3.50$\pm$0.04 & 0.86 \\
\hline
PKS 2155-304 & Uncorrected & 22.5 & std & 0.9$\pm$0.1 & 3.60$\pm$0.05 & 1.52 \\
PKS 2155-304 & Corrected & 22.5 & 22.5 & 2.5$\pm$0.1 & 3.47$\pm$0.04 & 1.70 \\
\hline
\end{tabular}
}
\end{center}
\caption{Parameters are given for a power law fit to the differential energy spectrum  $I(E)=I_{0}E^{-\Gamma}$, where $I_{0}$ is measured in photons cm$^{-2}$ s$^{-1}$ TeV$^{-1}$.  }
\label{tab2}
\end{table*}
\section{Conclusion}A  method for correcting for changes in low-level atmospheric quality is applied to the  variable source PKS 2155-304. The method, based on cosmic-ray trigger-rate, and Ceilometer input, has allowed a corrected set of fluxes for PKS 2155-304 to be produced from data that would otherwise be unusable. This is particularly important as this data set forms part of a large multi-wavelength campaign so removing atmospheric biases is vital. To the lowest order, the effect on integral gamma-ray flux is seen to be proportional to the zenith- and time-corrected cosmic-ray trigger-rate. \newline
A  single scattering LIDAR installed in 2007 at the H.E.S.S. site operate at wavelengths closer to Cherenkov light and has a greater range, and will hopefully allow more straight forward correction. As has been shown, though, the comparison of real and simulated reconstructed shower depth under the application of different atmospheric models allows a coarse appreciation of atmospheric conditions, which is a useful check for the more accurate LIDAR dataset expected to be obtained soon.  A more detailed examination of this technique is presented in \cite{bib:nolan}.

\par{\noindent {\it Acknowledgements}  The authors would like to acknowledge the support of their host institutions, and in addition we are grateful to Werner Hofmann, Michael Punch, Paula Chadwick \& Michael Daniel for their helpful comments. S.N. and C.R. acknowledge support from the Science and Technology Facilities Council of the UK. G.P. acknowledges support by the German Ministry for Education and Research (BMBF) through DESY grant 05CH5VH1/0 and DLR grant 50OR0502.}

\end{document}